\documentclass[twocolumn]{jpsj3}
\usepackage{txfonts}
\usepackage{graphicx,amsmath,xspace}
\usepackage{color}

\title{
Destruction of Magnetic Long-Range Order
by Hole-Induced Skyrmions in Two-Dimensional Heisenberg Model
}

\author{
Tomoharu Suda
\thanks{E-mail: suda.tomoharu.88s@st.kyoto-u.ac.jp}
and
Takao Morinari
\thanks{E-mail: morinari.takao.5s@kyoto-u.ac.jp}
}

\inst{
Graduate School of Human and Environmental Studies, Kyoto University, 
Kyoto 606-8501, Japan
}

\newcommand{\suda}[1]{\textcolor{black}{#1}}
\newcommand{\tm}[1]{\textcolor{black}{#1}}

\newcommand{\be}{\begin{equation}}
\newcommand{\ee}{\end{equation}}
\newcommand{\bea}{\begin{eqnarray}}
\newcommand{\eea}{\end{eqnarray}}

\date{\today}

\abst{
Motivated by the rapid destruction of antiferromagnetic long-range 
order in hole-doped cuprate 
high-temperature superconductors,
we study the effect of skyrmions on the magnetic long-range
order (MLRO).
Here we assume that either a skyrmion or antiskyrmion is introduced 
by a doped hole.
Our numerical simulation indicates that 
in the case of isolated skyrmions, there is 
an abrupt disappearance of MLRO
for doping concentration $x < 1.0\times 10^{-4}$.
In the case of skyrmion-antiskyrmion pairs,
the critical doping concentration $x_c$ for the suppression of MLRO 
is given as a function of the separation of the pairs.
For a moderate separation of $3 - 4$ lattice constants, 
we find that the critical doping is consistent 
with the experimental value.
}

\begin{document}

\maketitle

\section{Introduction}
The parent compounds of the cuprate high-temperature superconductors
are Mott insulators.\cite{LeeNagaosaWen2006}
Each hole at the copper sites is localized 
by the charge-transfer gap.\cite{Zaanen1985}
These localized spin-1/2 moments are well described by
the two-dimensional antiferromagnetic (AF) Heisenberg model, 
and it is now well established that the ground state 
is the antiferromagnetic 
long-range-ordered (AFLRO) state.\cite{Manousakis1991}
High-temperature superconductivity takes place upon hole doping
in such parent compounds.

In order to uncover the mechanism of high-temperature superconductivity
and the physics underlying 
the intriguing pseudogap phenomenon\cite{Keimer2015}
observed above the superconducting transition temperature,
it is necessary to figure out how doped holes are described
in the cuprate high-temperature superconductors.
In slightly hole doped compounds, the AFLRO state
is rapidly destroyed by hole doping.\cite{Luke1990}

Timm and Bennemann\cite{Timm2000} studied the effect of vortices 
on the AFLRO state.
They considered vortices induced by doped holes
as well as thermally excited vortices.
The N{\' e}el temperature was computed using extended 
Berezinskii--Kosterlitz--Thouless renormalization group theory
and good agreement with the experimentally observed
critical doping $x_c \simeq 0.02$ was obtained.

In this paper, we consider skyrmions instead of vortices.
It is natural to consider that vortices destroy magnetic long-range 
order (MLRO) in the XY model.
The field in the XY model is the U(1) phase field.
Thus, a topological defect in such a U(1) theory is a vortex.
On the other hand, the low-energy effective theory
for the AF Heisenberg model is the O(3) nonlinear
$\sigma$-model.\cite{Chakravarty1988,Chakravarty89}
A topological defect in theories with O(3) symmetry 
is a skyrmion configuration.
If we denote the angle of the XY spin by $\phi$,
a single vortex solution existing at the origin is
$\phi  = {\tan ^{ - 1}}\left( y/x \right)$.
This vortex is interpreted as a gauge flux with
the vector potential
\begin{equation}
 {\partial_x} \phi  =  - y/{r^2},\hspace{2em}
  {\partial_y} \phi  = x/{r^2}.
  \label{eq_v}
\end{equation}
This flux is singular at the core of the vortex,
and the winding number is one for vortices.
Thus, this is a topological defect.
In the case of the Heisenberg model 
skyrmions are exact solutions of 
the O(3) nonlinear $\sigma$ model \cite{Belavin1975}
that describes the low-energy effective theory 
of the AF Heisenberg model.\cite{Chakravarty1988,Chakravarty89}
In the CP$^1$ representation \cite{Rajaraman} of the 
nonlinear $\sigma$ model,
such a skyrmion solution turns out to be the gauge flux
$\partial_x a_y - \partial_y a_x$
with 
\begin{equation}
{a_x} =  - \frac{y}{{{r^2} + {\lambda ^2}}},\hspace{2em}
{a_y} = \frac{x}{{{r^2} + {\lambda ^2}}}.
\label{eq_a}
\end{equation}
Here $\lambda$ is the size of the skyrmion.
The skyrmion configuration is a topological defect
characterized by the homotopy group $\pi_2(S^2)\simeq \mathbb{Z}$.
Note that Eq.~(\ref{eq_a}) coincides with
Eq.~(\ref{eq_v}) when taking the limit $\lambda \rightarrow 0$.
In contrast to vortices in the XY model,
the singularities at the positions of the vortices
are relaxed by a finite parameter $\lambda$.
For these reasons, it is natural to consider skyrmions
instead of vortices in the Heisenberg model.
However, it has not yet been established how 
a doped hole creates a skyrmion microscopically.
Numerical study of the Hubbard model within 
the unrestricted Hartree--Fock approximation\cite{Verges1991}
suggests the existence of many nearly degenerate
metastable configurations and 
that a vortexlike spin configuration is formed around the holes.
An unrestricted spin-rotational-invariant slave-boson 
approach\cite{Seibold1998}
suggests that a vortex-antivortex pair has lower energy
than the N\'{e}el-type bipolaron for two holes doped 
in the half-filled Hubbard model.
An effective Hamiltonian approach based 
on the $t$-$J$ model\cite{Shraiman1988,Shraiman90}
suggests that a long-range dipolar distortion of 
the staggered magnetization is created around doped holes.
Skyrmions have been investigated using
the analogies between pions in QCD and magnons 
in antiferromagnets.\cite{Bar2004,Wiese2005}
A finite-size-cluster calculation suggets
the stability of a skyrmion around 
a localized doped hole.\cite{Morinari2012,Haas1996,Gooding91}
Experimentally, the existence of skyrmions in AF insulators 
has been supported by the low-temperature magnetic and transport 
properties of slightly Li-doped La$_2$CuO$_4$.\cite{Raicevic2011}

In this paper we numerically simulate the destruction 
of the AFLRO state assuming that doped holes induce skyrmions.
We examine two cases:
one is that isolated skyrmions are introduced 
by doped holes;
the other is that skyrmion-antiskyrmion pairs 
are introduced by doped holes.
We find that isolated skyrmions are too strong to destroy 
the AFLRO state.
\tm{
The same is true for skyrmion-skyrmion pairs.
}
If we consider a skyrmion-antiskyrmion pair
with a moderate separation, the critical doping $x_c$ is 
in good agreement with the experimental value.\cite{Luke1990}

The outline of this paper is as follows.
In Sect.~\ref{sec:model} we describe the model
and the numerical simulation procedure.
In Sect.~\ref{sec:result} we present the numerical 
simulation results.
Finally Sect.~\ref{sec:conclusion} is devoted to
a conclusion.

\section{Model}
\label{sec:model}
In order to investigate the effect of skyrmions on the AFLRO state,
we randomly distribute skyrmions or antiskyrmions in the spin system.
For comparison, we also investigate the effect of vortices on 
magnetic long-range order in the XY model.
In that case, we randomly distribute vortices or antivortices 
in the spin system.

\tm{
Here we focus on the doping concentration
at which the system is insulating for the cuprates.
Thus, we neglect the effect of itinerant electrons.
There is a possibility that the doped-hole wave function
is not restricted to a single site but is spread over
several lattice sites.
Although such a wave function might lead to the stabilization 
of skyrmions,\cite{Morinari2012} 
we do not include it to simplify the
numerical simulation.
}

For the distribution of skyrmions,
we consider two cases.
One is a completely random distribution
of skyrmions and antiskyrmions
with the constraint that the number of skyrmions
is equal to the number of antiskyrmions.
The other is a random distribution
of skyrmion-antiskyrmion pairs.
We consider the same distributions 
for the case of vortices in the XY model.

Since we are interested in the spin system with MLRO,
we may consider classical spins.
The effect of quantum fluctuations can be included 
by reducing the size of spins, which leads to
renormalization of the exchange interaction $J$ between spins.
We take a square lattice with a size of $N \times N$,
and the Hamiltonian is given by
\begin{equation}
 H = J\sum\limits_{\left\langle {i,j} \right\rangle } {{{\bf{S}}_i}
  \cdot {{\bf{S}}_j}},
\end{equation}
where ${\bf S}_j$ is the localized spin at site $j$
and the summation is taken over the nearest-neighbor sites.
In the case of XY spins, ${\bf S}_j$ is a two-dimensional vector,
${\bf S}_j = S(\cos \phi_j, \sin \phi_j)$
with $S$ the size of the spin.
In the case of Heisenberg spins, ${\bf S}_j$ is a three-dimensional vector,
${\bf S}_j 
= S(\sin \theta_j \cos \phi_j, \sin \theta_j \sin \phi_j, \cos \theta_j)$.
For classical spins, one can transform antiferromagnets 
to ferromagnets 
on bipartite lattices
by reversing the direction of spins 
at odd sites.
Thus, we may consider a ferromagnet instead of an antiferromagnet.

We assume that skyrmions or vortices are induced by doped holes.
We denote the number of doped holes by $N_h$.
The doped hole concentration is defined by $x=N_h/N^2$.
In the case of vortices in the XY model, 
the randomly distributed $N_h$ vortices
are described by
\begin{equation}
 \phi_j = \sum\limits_{\ell  = 1}^{{N_h}} {{q_\ell
  }{{\tan }^{ - 1}}\left( {\frac{{y_j - {y_\ell }}}{{x_j - {x_\ell }}}}
		   \right)},
\label{eq_vortices}
\end{equation}
where $x_j$ and $y_j$ denote the coordinate of lattice site $j$
and $q_{\ell} = 1$ for vortices and $q_{\ell} = -1$ for antivortices. 
The vortex positions are denoted by $(x_{\ell},y_{\ell})$,
and random integer numbers ranging from 1 to $N$
plus one-half are assigned to ${x_\ell }$ and ${y_\ell }$.
Thus, vortices are introduced at the centers of plaquettes.

In the case of skyrmions in the Heisenberg model, 
we make use of the form of Eq.~(\ref{eq_vortices}).
We fix the size of the skyrmion $\lambda$
to the lattice constant.\cite{Morinari2012}
We determine the polar angle $\theta_j$ 
of the Heisenberg spin,
which is absent in the XY spin,
through the minimization of the energy
of the spin system.
We realign each spin so that it is in the direction of 
the local average of the four surrounding spins.
We update the spin configuration 
while keeping the directions of 
the four spins surrounding the skyrmion or antiskyrmion fixed.
\tm{
This is necessary to maintain the skyrmion configurations
because skyrmions are metastable in the Heisenberg model.
In order to stabilize skyrmions,
we require additional interactions, such as the Dzyaloshinskii--Moriya 
interaction or frustrated interaction.
However, the mechanism by which skyrmions in the cuprates
are stabilized
has not been established yet, 
and so we simply introduce the fixing procedure
above to maintain the skyrmions.
}
We carry out the update procedure until the self-consistent
spin configuration is obtained.


\section{Numerical Simulation of MLRO Destruction}
\label{sec:result}
In order to measure the effect of MLRO destruction 
by skyrmions, we analyze the skyrmion concentration dependence 
of the magnetization,
$M(N) = \sqrt { \left( \sum_{i=1}^{N^2}\ {\bf S}_{i} \right)
^{2}}/(N^2S)$,
which is normalized by the maximum magnetization, $N^2S$.
The thermodynamic limit of the magnetization,
${M_0} = \mathop {\lim }\limits_{N \to \infty } M\left( N
\right)$,
is obtained from finite-size scaling 
assuming that
$M\left( N \right) = {M_0} + {M_1}/N^2 + {M_2}/{N^4} + \cdots$.
We carried out the numerical simulation for $N=128, 256, 512$.

A simple random configuration is obtained by distributing skyrmions 
and antiskyrmions completely randomly
\suda{by choosing $N_h$ sites 
out of $N^2$ sites}. 
However, this completely random configuration leads to
the abrupt disappearance of MLRO.
The critical value of $x$ is less than $10^{-4}$.
The situation is almost the same for the case of vortices
in the XY model.
Note that here we set $\lambda$ equal to the lattice 
constant.
This choice corresponds to the
half-skyrmion.\cite{Morinari2005,MorinariMFS}
\tm{
In the CuO$_2$ plane there is a strong correlation of 
stabilizing the Zhang--Rice singlet.\cite{ZhangRice1988}
If a doped hole forms the Zhang--Rice singlet with a copper site spin,
then the spin is removed at the doped hole position.
The singlet state tends to stabilize a half-skyrmion-like spin configuration
as discussed in Ref.~\citen{Morinari2012}.
}
For larger values of $\lambda$, the disorder introduced by skyrmions
becomes stronger.
Meanwhile, the case of $\lambda=0$ must be considered separately.
In this case, only local spin flips are introduced,
and the magnetization immediately recovers the bulk value
away from holes.
Thus, we simply obtain $M=1-2x$, which is inconsistent 
with the rapid destruction
of MLRO in the cuprates.

\begin{figure}
   \begin{center}
    \includegraphics[width=0.8 \linewidth]{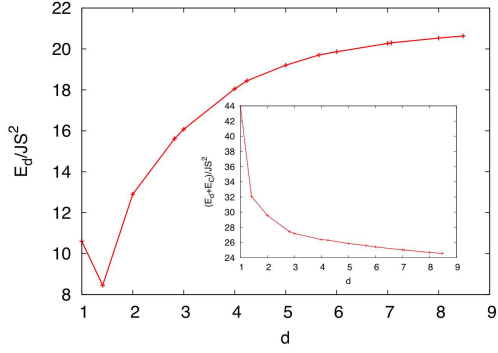}
   \end{center}
   \caption{ 
 \label{fig:dipole_energy}
 (Color online)
 Single skyrmion-antiskyrmion pair 
 energy $E_{\rm d}$ as a function of $d$.
 The energy is normalized by the spin system energy $JS^2$.
 The inset shows the pair energy with the Coulomb energy between
 the holes,
 $E_{\rm C}={e^2}/\left( {4\pi \varepsilon d} \right)$.
 (The solid lines are a guide to the eye.)
 Here we set the dielectric constant $\varepsilon=3\varepsilon_0$ 
 with $\varepsilon_0$ the dielectric constant in the vacuum
 and $JS^2=330$ K.
 $E_{\rm d}+E_{\rm C}$ is a monotonically decreasing function
 with respect to $d$.
 The behavior is qualitatively similar for moderate changes
 in $\varepsilon$ and $JS^2$.
 }
 \end{figure}

A completely random configuration ignores
the interaction effect between skyrmions
as well as the Coulomb interaction between holes.
If we denote the separation of a skyrmion-antiskyrmion pair
by $d$,
the energy of a single pair depends on $d$.
Figure \ref{fig:dipole_energy} shows the energy $E_{\rm d}$
of a skyrmion-antiskyrmion pair
as a function of $d$.
We computed the energy of a single pair whose direction
is either horizontal or diagonal.
For $d>3$, the energy does not depend on the direction
of the pair.
The energy saturates for large $d$, 
which can be interpreted as 
meaning
that the pair is almost decoupled and 
the configuration of pairs
cannot be distinguished from completely randomly distributed skyrmions.
Meanwhile, for $d<3$, the result slightly 
depends on the direction of the pair.
The effect of the Coulomb energy $E_{\rm C}$ between
doped holes is shown in the inset of Fig.~\ref{fig:dipole_energy}.
Note that $E_{\rm d}+E_{\rm C}$ is a monotonically decreasing
function with respect to $d$.
On the other hand, the average distance between the doped holes
is given by $1/\sqrt{x}$ in units of the lattice constant.
For the case of $x=0.02$, we find that $1/\sqrt{x} \simeq 7.1$.
The value of $1/\sqrt{x}$ gives an upper bound for $d$.
Thus, we may consider $1<d<7$ when examining the effect 
of skyrmion-antiskyrmion pairs on MRLO.

\begin{figure}
      \begin{center}
       \includegraphics[width=0.8 \linewidth]{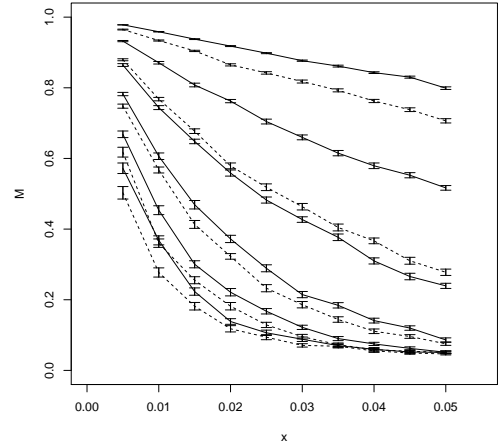}
      \end{center}
      \caption{ 
      \label{fig:magnetization_skyrmion_d}
      Magnetization $M$ as a function of 
      doped-hole concentration $x$
      for the Heisenberg model.
      $M$ is normalized by $N^2S$.
      Solid lines are for the case of 
      skyrmion-antiskyrmion pairs aligned horizontally.
      The values of $d$ are
      $d = 1,2,3,4,5,6$ from top to bottom.
      Dashed lines are for the case of 
      skyrmion-antiskyrmion pairs aligned diagonally.
      The values of $d$ are
      $d = \sqrt{2},2\sqrt{2},3\sqrt{2},4\sqrt{2},5\sqrt{2}$ 
      from top to bottom. 
      (The solid lines and dashed lines are a guide to the eye.)
      }
     \end{figure}

We carried out numerical simulations for different values of $d$.
In units of the lattice constant,
we take $d=1,2,3,4,5,6$ for pairs aligned horizontally
and $d=\sqrt{2},2\sqrt{2},3\sqrt{2},4\sqrt{2},5\sqrt{2}$ for 
pairs aligned diagonally.
\suda{
The configurations of the pairs are determined randomly. 
In order to ensure that the  separation of the pairs 
is the prescribed value $d$, we first choose $N_h/2$ consecutive 
sites of size $d+1$ without overlapping and then place
skyrmion-antiskyrmion pairs at both their ends.
The number of samplings is 20, which appears to be sufficient as 
we did not observe significant improvement of the convergence 
by increasing the number of samplings.}
We found systematic variation of the $x$-dependence of the magnetization 
upon increasing $d$ as shown in Fig.~\ref{fig:magnetization_skyrmion_d}.
For $d\leq 3$, 
which is in the  parameter range where the single pair energy 
depends on the pair direction,
$M$ decreases slowly and 
almost linearly with respect to $x$.

This gradual destruction of MLRO by skyrmion-antiskyrmion pairs changes
to rapid destruction for $d\geq 4$
as shown by the convex downward curves.

\begin{figure}
	\begin{center}
	 \includegraphics[width=0.8 \linewidth]{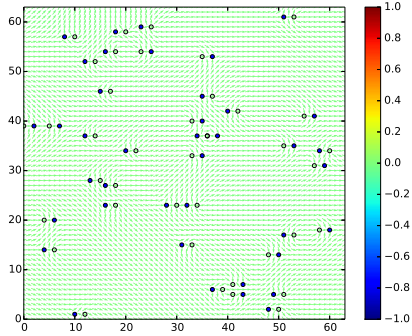}
	\end{center}
	\caption{ 
	\label{fig:spin_skyrmion_small}
	(Color online)
	Spin configuration of the Heisenberg spins with 
	skyrmion-antiskyrmion pairs with $d=2$. Skyrmions and
	antiskyrmions are denoted by filled and open circles, respectively.
	The doping concentration is $x=0.02$.
	The $x$ and $y$ components of spins are denoted by arrows
	and the $z$ component is shown in the color scale.
	}
       \end{figure}

\begin{figure}
	 \begin{center}
          \includegraphics[width=0.8 \linewidth]{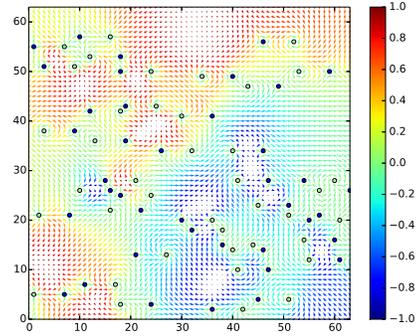}
	  \includegraphics[width=0.8 \linewidth]{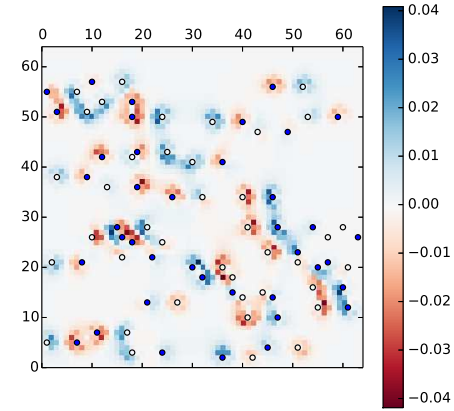}
	 \end{center}
	 \caption{ 
	 \label{fig:spin_skyrmion_large}
	 (Color online)
	 Spin configuration (upper panel)
	 and topological charge density distribution (lower panel)
	 of the Heisenberg spins with skyrmion-antiskyrmion pairs
	 for the case of $d = 6$ with $x=0.02$.
	 Skyrmions and antiskyrmions are denoted by filled and open
	 circles, respectively.
	 }
	\end{figure}

The difference between these two cases is clarified 
when we see the spin configuration and topological charge
distribution.
The topological charge density on the lattice is defined by
\begin{eqnarray}
q\left( {{x_j},{y_j}} \right) 
&=& \frac{1}{{16\pi }}{\bf{n}}\left({{x_j},{y_j}} \right) 
\cdot \left[ {{\bf{n}}\left( {{x_j} + 1,{y_j}} \right) \times
  {\bf{n}}\left( {{x_j},{y_j} + 1} \right)} \right. \nonumber \\
& & \left. + {\bf{n}}\left( {{x_j} - 1,{y_j}} \right) \times {\bf{n}}\left(
	       {{x_j},{y_j} - 1} \right) \right. \nonumber \\
& & \left. + {\bf{n}}\left( {{x_j},{y_j} + 1}
						   \right) \times
	       {\bf{n}}\left( {{x_j} - 1,{y_j}} \right) \right. \nonumber \\
 & & \left. { + {\bf{n}}\left( {{x_j},{y_j} - 1} \right) \times
      {\bf{n}}\left( {{x_j} + 1,{y_j}} \right)} \right],
\end{eqnarray}
where ${\bf n}(x_j,y_j)={\bf S}_j/S$.
Figure \ref{fig:spin_skyrmion_small} shows
the spin configuration for the case of $d=2$.
In spite of the fact that we introduced skyrmions,
almost all the spins lie in the $x$-$y$ plane.
Reflecting this feature, there is almost no 
topological charge density distribution over the system.
In contrast, in the case of $d=6$,
both the spin configuration and the topological charge density 
distribution shown in Fig.~\ref{fig:spin_skyrmion_large}
suggest that there is a significant effect
of skyrmion-antiskyrmion pairs on the spin configuration.
\tm{
Although the topological charge density is not directly
connected with the disorder of spins, 
as one can see from 
Fig.~\ref{fig:spin_skyrmion_large},
the rapid change in
the sign of the topological charge density over the sites implies
the presence of strong disorder.
}

Now we discuss the critical doping concentration.
We stop the update procedure when the maximum difference between
the spin values after the update and those before
the update is less than $10^{-6}$.
Thus, for the size of $512 \times 512$, the value of $M\sim 0.1$
corresponds to the complete destruction of MLRO.
Therefore, we may conclude that MLRO should disappear
around $0.02 < x < 0.04$ 
\tm{for $d > 3$.}
From the discussion above, the Coulomb repulsion 
increases $d$, the separation of skyrmion-antiskyrmion pairs,
while there is an upper limit of $1/\sqrt{x}$ for $d$.
Thus, the rapid destruction of MLRO by skyrmion-antiskyrmion pairs 
is consistent with that observed in the cuprates.

\begin{figure}
      \begin{center}
       \includegraphics[width=0.8 \linewidth]{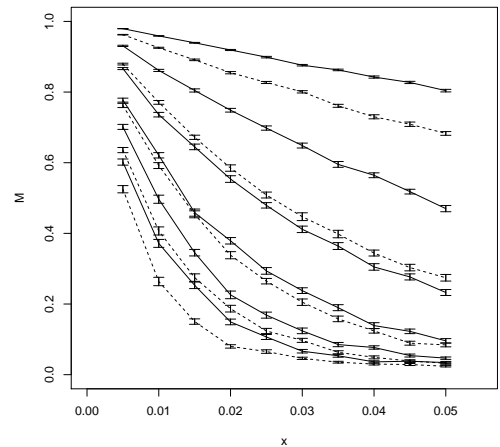}
      \end{center}
      \caption{ 
      \label{fig:magnetization_vortex_d}
      Magnetization $M$ as a function of 
      doped-hole concentration $x$
      for the XY model.
      Solid lines are for the case of 
      vortex-antivortex pairs aligned horizontally.
      The values of $d$ are
      $d = 1,2,3,4,5,6$ from top to bottom.  
      Dashed lines are for the case of 
      vortex-antivortex pairs aligned diagonally.
      The values of $d$ are
      $d = \sqrt{2},2\sqrt{2},3\sqrt{2},4\sqrt{2},5\sqrt{2}$
      from top to bottom.  
      (The solid lines and dashed lines are a guide to the eye.)
      }
     \end{figure}

For comparison, we carried out similar numerical simulations
for the case of 
\tm{vortex-antivortex pairs}
introduced in the XY model.
The result is shown in Fig.~\ref{fig:magnetization_vortex_d}.
Interestingly, we obtained a similar result.

\begin{figure}
      \begin{center}
       \includegraphics[width=0.8 \linewidth]{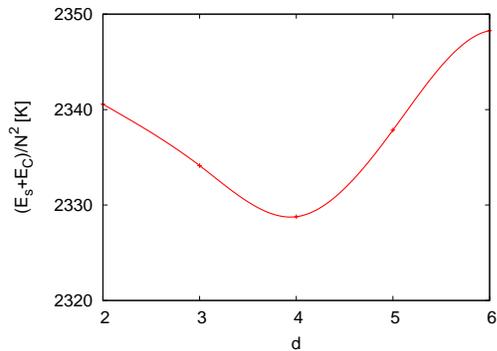}
      \end{center}
      \caption{ 
      \label{fig:Eallx0_02}
      (Color online)
      System energy versus $d$ for $x=0.02$.
      The energy was computed for a size of 256 $\times$ 256
      with the number of samplings being 20.
      The minimum is around $d\sim 4$ as a consequence
      of the competition between the energy of the spins, $E_s$,
      and the Coulomb interaction energy between the holes, $E_C$.
      (The solid line is a guide to the eye.)
      }
     \end{figure}

\tm{
In order to find the critical value of $d$,
we computed the energy of the system.
Figure \ref{fig:Eallx0_02} shows 
the $d$ dependence of the system energy
for $x=0.02$,
which consists of the energy of the spins 
and the Coulomb interaction energy between the holes.
The system energy shows a minimum around $d\sim 4$.
The appearance of this minimum is a natural consequence
of the competition between the energy of the spins and the 
Coulomb interaction energy between the holes.
Basically, the energy of the spins increases monotonically 
with increasing $d$.
Meanwhile, the Coulomb interaction energy decreases
monotonically with increasing $d$.
This is understood from an approximate calculation.
An approximate formula for the Coulomb energy is given by
\begin{equation}
{E_C}\left( d \right) \simeq \frac{{{e^2}}}{{4\pi \varepsilon }}
 \left({\frac{{{N^2}x}}{{2d}} + \frac{3}{2}{N^3}{x^2}} \right).
 \label{eqECapp}
\end{equation}
This formula is obtained by finding an approximate formula
for the Coulomb energy of two skyrmion-antiskyrmion pairs,
and then computing the total energy of randomly distributed
skyrmion-antiskyrmion pairs.
The term proportional to $N^3$ on the right-hand side 
of Eq.~(\ref{eqECapp}) is the dominant term.
For the case of $x=0.02$ and $N=256$, we find that
its value is 2250 K per site.
This is consistent with Fig.~\ref{fig:Eallx0_02}.
However, this value is $d$-independent.
We confirmed that the energy takes a minimum
around $d \sim 4$ for different values of $x$
ranging from 0.01 to 0.05.
The critical value of $d$ depends on 
$\eta  = {e^2}/\left( {4\pi \varepsilon aJ{S^2}} \right)$
with $a$ the lattice constant.
For the cuprates, $\eta=44.4$ with $\varepsilon=3\varepsilon_0$ and $S=1/2$.
However, the critical value of $d$ does not change much
upon changing $\eta$ from 30 to 120.
Thus, even if there is a reduction of the size of the spin
due to the quantum fluctuations, the critical value of $d$
is the same.
Note that the statistical error for the energy calculation
is significant.
Thus, we were unable to evaluate the binding energy of
the skyrmion-antiskyrmion pair.
}

 \section{Conclusions}
     \label{sec:conclusion}
We have examined 
the effect of skyrmions on MLRO
in the Heisenberg model.
We have assumed that the size of the skyrmion, $\lambda$,
is equal to the lattice constant.
Either a skyrmion or antiskyrmion 
is introduced by doping holes in the system.
We found that the effect of isolated skyrmions and antiskyrmions
is too strong to destroy MLRO.
For the case of skyrmion-antiskyrmion pairs,
which are expected to be formed because of the Coulomb repulsion
between holes and the interaction between skyrmions,
we found that
the critical hole doping concentration 
is consistent with the experimental values for the cuprates.
Since the appropriate pair separation 
is almost equal to the average hole distance $1/\sqrt{x}$,
our numerical simulations suggest that
doped holes are almost uniformly distributed
and that there is an antiferromagnetic configuration
of skyrmions and antiskyrmions.

	{\footnotesize \section*{Acknowledgements} 
	We thank Kenji Harada and Kenji Kubo for helpful discussions.
	}

\bibliography{../../../references/tm_library2}

\end{document}